\documentclass{article}
\usepackage{graphicx}
\usepackage[english]{babel}
\usepackage{srcltx}
\usepackage{amsmath}
\usepackage{amssymb}
\usepackage{amsfonts}
\usepackage{latexsym}
\usepackage{textcomp}
\usepackage{appendix}
\usepackage{multirow}
\usepackage{booktabs}
\usepackage{rotating}
\usepackage{url}
\usepackage{array}
\usepackage{marvosym}
\usepackage{afterpage}
\usepackage{pdflscape}
\usepackage[table,svgnames]{xcolor}
\usepackage{graphics}
\usepackage{graphicx}
\usepackage{caption}
\usepackage{multirow}
\usepackage{booktabs}
\usepackage{standalone}
\usepackage{tabulary}
\usepackage{adjustbox}
\usepackage{booktabs}
\usepackage{pdflscape,afterpage,caption}
\usepackage{longtable}
\usepackage[round,authoryear]{natbib}
\usepackage{subfig} 
\usepackage{lineno,hyperref}
\modulolinenumbers[5]
\usepackage[margin=1.25in]{geometry}

\title{One model is not enough: heterogeneity in cryptocurrencies' multifractal profiles}

\author{Aurelio F. Bariviera\thanks{Corresponding author: aurelio.fernandez@urv.cat} \\
 \scriptsize Universitat Rovira i Virgili, Department of Business, Av. Universitat 1, 43204 Reus, Spain}

\begin{document}
\maketitle

\begin{abstract}
This letter studies of the multifractal dynamics in 84 cryptocurrencies. It fills an important gap in the literature, by studying this market using two alternative multi-scaling methodologies. We find compelling evidence that cryptocurrencies have different degree of long range dependence, and --more importantly -- follow different stochastic processes. Some of them follow models closer to monofractal fractional Gaussian noises, while others exhibit complex multifractal dynamics. Regarding the source of multifractality, our results are mixed. Time series shuffling produces a reduction in the level of multifractality, but not enough to offset it. We find an association of kurtosis with multifractality.

\end{abstract}

\section{Introduction \label{sec:intro}}
Contemporary to the outbreak of the 2008 financial crisis, an anonymously posted paper attributed to \cite{Nakamoto}, set the grounds for a new type of financial asset. This new synthetic product, aimed at bypassing the traditional banking system, became known as cryptocurrency. In spite of the fact that its properties as ``currency'' has been cast in doubt by \cite{NBERw19747}, \cite{DWYER201581}, \cite{SELGIN201592}, and \cite{SCHILLING201916}, it is undoubtedly a financial asset of great interest among investors. Shortly after its launching, Bitcoin became tantamount of cryptocurrency. This success led many entrepreneurs to develop their own crytpocurrencies. \cite{Elbahrawy2017} trace the evolutionary dynamics of this market, finding that until the beginning of 2017 the average birth rate of new cryptocurrencies was slightly larger than the average death rates, with an average net increment in the number of coins in the long run. 
As of March 2020, there are more than 5000 cryptocurrencies, which are traded on 20877 platforms, adding up a market capitalization of 142 USD billions \citep{coinmarketcap}. These figures highlight the economic relevance of this phenomenon. 

Cryptocurrencies studies emerge as new and frutiful empirical area, where researchers look for insights of this novel product. Recent surveys \citep{Yli-Huumo2016,Corbet2019,MeredizBariviera2019} show aspects that has been covered until now: statistical properties of daily returns \citep{Urquhart2016,Barivieraetal2017}; safe haven characteristics of Bitcoin \citep{BOURI2017192,Smales2018}; correlation of main cryptocurrencies with traditional assets \citep{CORBET201828,Aslanidis2019}; and  portfolio optimization \citep{PLATANAKIS2019}.

Nevertheless, there are several gaps in the literature. Firstly, most empirical studies focus their attention on Bitcoin, or at most on the five biggest cryptocurrencies (Bitcoin, Ethereum, Bitcoin Cash, Ripple, Litecoin). Secondly, the scale dimension remains unstudied.

Hence, an alternative approach is necessary.
This letter expands and complements previous literature on cryptocurrencies in three aspects: (i) it uses a multi-scaling approach that represents a new approach to this market; (ii) it works with a comprehensive set of cryptocurrencies, which reflects more accurately the behavior of the market; and (iii) it presents a criterion for selecting more appropriate stochastic models of cryptocurrencies dynamics.

The letter is organized as follows. Section \ref{sec:methods} discusses general aspects of long range memory and explains the methodology. Section \ref{sec:data} presents the data. Section \ref{sec:results} discusses the main findings. Finally, Section  \ref{sec:conclusions} outlines the conclusion of our analysis.

\section{Methods \label{sec:methods}}
\subsection{Long range memory}
The Efficient Market Hypothesis (EMH) represents the cornerstone of financial economics, upon which many areas (e.g. portfolio optimization, option pricing) are built.
 It is based on the idea price movements in a competitive market constitute a fair game. The 
pioneering work by  \citet{Bachel} proposes the the arithmetic Brownian motion as a suitable model to represent the price dynamics of French bonds. 
Several decades later, \citet{Samuelson65} rediscovers Bachelier's model, proposing a geometric Brownian motion model\footnote{The geometric Brownian motion had been independently proposed by \cite{Osborne1959,Osborne1962}} and sets the grounds for the EMH. 
The definition and classification of the EMH is owed to \citet{Fama70}. Briefly, the EMH requires that returns of financial assets follow a Markov process with respect to the respective underlying information set. 

Contrary to this ideal model, several papers find long memory in traditional financial assets, using different methods \citep{BarkoulasBaumTravlos00,Carbone04,McCarthy09,CajueiroGogasTabak09}. An important research line in statistics and econometrics is directed at detecting long memory in financial time series. 
Different alternatives have been formulated (e.g.: fractional Brownian motion, fractional L\`evy flights) to account for long memory. However, they are all defined within a monofractal framework. 
Even Autoregressive Fractionally Integrated Moving Average (ARFIMA) processes, proposed by \cite{GrangerJoyeux1980} as a generalization of \cite{BoxJenkins} models, assume a monofractal stationary process.

Regarding cryptocurrencies, there are also several studies on long range dependence. \cite{Bariviera2017} shows a declining trend in long range dependence in Bitcoin returns, but a persistent long range memory in daily volatility. \cite{TIWARI2018106} observe that the Bitcoin market has a trend towards the informational efficiency, albeit it exhibits long range dependence during April-August, 2013 and August-November, 2016. Such results are aligned with those previously found by \cite{Urquhart2016}. More recently, \cite{PHILLIP201995} show that faster transacted currencies show stronger oscillating long run autocorrelations. \cite{Kristoufek2019} examine and produce an efficiency ranking of fourteen coins and tokens according to the well-established Efficiency Index developed by \cite{Kristoufek2013}. For the sake of brevity, we refer to \cite{Corbet2019} and \cite{MeredizBariviera2019} for further details on the empirical financial literature on cryptocurrencies.  

In a recent contribution, \cite{Kukacka2020} link the concepts of multifractality and complexity, and replicate the statistical dynamic properties of the time series by means of several agent-based models. In a broad sense, we can say that multifractality is a statistical feature of time series realated to its non-trivial scaling properties. For a comprehensive discussion on multifractality in financial markets see the recent review by \cite{Jiang2019}

\subsection{Generalized Hurst exponent}
Harold Edwin Hurst, a British engineer, developed in a series of papers \citep{Hurst1951,Hurst1956a,Hurst1956b,Hurst1957} an empirical method to measure the long range dependence of river discharges. His method marked a landmark in hydrology studies, and was found to have applications in other scientific domains. The original method, R/S, was based on the rescaled range of the partial sums of deviations of a time series from its mean.  

\cite{MandelbrotVanNess1968} postulate a generalization Brownian motion and Gaussian noise models, allowing for long range dependence, linked to the system's Hurst exponent. \cite{Mandel68} find that these fractional models behave remarkably good in hydrology, and  \cite{Mandel72} proposes its use in economics. 

Depending on the type of signal under analysis, and the goal of the research, we can select among a wide range of methods to compute the Hurst exponent. An exhaustive discussion on this topic is in  \citet{Serinaldi10}. 

The Hurst exponent $H$ does not only measure long range dependence, but it is also closely related to the fractal dimension of a time series, as shown by \cite{Sanchez-Granero2012}. In economics, it is common to assume that the process under study is a fractional Brownian motion (fBm) or a fractional Gaussian noise (fGn). They are monofractal processes, meaning that the Hurst exponent scales linearly. However, real world phenomena can exhibit more complex dynamics. For example,  \cite{TUBERQUIADAVID2016373} show that network traffic pattern present multifractal characteristics, meaning that the Hurst exponent scales nonlinearly.

In this paper we use two approaches to detect multifractality. The first one is the generalized Hurst exponent (GHE) developed in \cite{DiMatteo2003}, and the second one is a multifractal version of the detrended fluctuation analysis presented in \cite{Kantelhardt2002}. It was recently recognized \citep{LuxSegnon2018,Buonocore2020} that multifractality could be considered an stylized fact of financial time series complementing those originally proposed by \cite{Cont}.

\subsection{The GHE estimator}
In contrast to other methods, this first method is specially suitable for describing the multi-scaling properties in financial time series. \cite{DiMatteo2003,DiMatteo2005} show that it provides robust and unbiased estimators on long term memory. 

Without loss of generality, let consider a financial time series $X(t)$ (with $t=\nu, 2\nu,\dots,k\nu,\dots, T$).  We are interested in analyzing the $q-$order moments of the distributions of increments, according to the time resolution ($\nu$). \cite{DiMatteo2007} reveals that $q$th-order moments are much less sensitive to outliers, and are associated with different features of the multi-scaling complexity of the time series. It is defined as
\begin{equation}
K_q(\tau)=\frac{\langle |X(t+\tau)-X(t)|^q\rangle}{\langle |H(t)^q|\rangle}
\end{equation}
where $\langle \cdot \rangle$ is the expectation operator. The generalized Hurst exponent $H(q)$ results from the scaling behavior  of $K_q(\tau)$ from the following relation:
\begin{equation}
K_q(\tau) \sim \left(\frac{\tau}{\nu}\right)^{q X(q)}
\end{equation}

This approach enables to signalize two situations: (a) uniscaling or unifractal processes where $H(q)=H$ is constant; and (b) multi-scaling or multifractal processes where $H(q)$ depends on $q$.

This procedure raised some concerns by \cite{Kukacka2020}, who argue that sometimes GHE yields inconsistent results.

\subsection{Multifractal Detrended Fluctuation Analysis}

To counterbalance the possible drawbacks of the GHE estimator, we also utilize an alternative approach to the degree of multifractality. We employ the Multifractal Detrended Fluctuation Analysis (MF-DFA). This method, developed by \cite{Kantelhardt2002} is generalization of the celebrated Detrended Fluctuation Analysis by \cite{Mosaic94}. This is method is divided into five steps. The first step determines the profile function:
\begin{equation}
Y(i)=\sum_{k=1}^{i}(X_t-\langle X \rangle), \;  i=1,\dots,N
\end{equation}
The second step partitions $Y(i)$ into nonoverlapping segments of length $s$. Then the third step computes the local trend for each segment using least square fitting, using linear, cuadratic or higher order polynomials.
The fourth step  computes the average over all segments, in order to obtain the $q$th order fluctuation function: 
 \begin{equation}
F_q(s)=\left(\frac{1}{2N_s}\sum_{\nu=1}^{2N_s}(F^2_{DFA}(\nu,s))^{q/2}\right)^{1/q}
\end{equation}
Finally, the fifth step determine the scaling behavior drawing using the $q \times F_q(s)$ log-log plane
\begin{equation}
F_q(s) \sim s^{H(q)}
\end{equation}

\cite{Kukacka2020} proposes the use of the range of the generalized Hurst exponents ($\Delta H\equiv \max_q H(q)-\min_q H(q)$) and the width of the multifractal spectrum  ($\Delta \alpha\equiv \max_q \alpha(q)-\min_q \alpha(q)$) to measure the degree of multifractality in a time series.  

\cite{Ihlen2012} provides a guide to implement MF-DFA to time series in Matlab. Its code, available in \cite{IhlenSoft}, is used in this paper. 

\subsection{The source of multifractality \label{sec:sourceMF}}
It is not only important to detect the presence of multifractality in time series, but also to determine its source. According to \cite{Kantelhardt2002} two types of multifractality can be distinguished. The first type is related to the probability density function, and the second one is related to the varying long-range correlation structure in the time series. 

In order to detect which kind of multifractality presents a time series, some tests on surrogated time series should be done. An straightforward strategy is to shuffle the original time series. This randomization will destroy non trivial correlations present in the original time series. Consequently, this procedure will allow to determine if the source of multifractality is the correlation structure. However, if the time series present multifractality due to the probability distribution, the measures of multifractality will not (significantly) change. Additionally, as highlihted by \cite{Kantelhardt2002} if both types of multifractality affect the time series, multifractality metrics will show a reduction in the shuffled series. A detailed discussion on the sources of multifractality in financial assets could be found in \cite{Barunik2012}.

\section{Data \label{sec:data}}
Cryptocurrencies' markets are not regulated by national authorities and market data lacks of proper independent standardization and verification. Consequently, a careful selection of the data sources is a key element in order to obtain reliable results.

Following \cite{Alexander2020}, we obtain our data from \cite{cryptocompare}, because other coin-ranking sites base their quotes on unreliable volume data. 

We use daily price data of the eighty-four largest cryptocurrencies (coins and tokens), according to traded volume. The period under examination goes from 06/01/2018 to 05/03/2020, for a total of 790 observations. The selection criteria was based on the average daily volume traded over the period, and the availability of data for every day within the period under study.

A table with the list and descriptive statistics of daily logarithmic returns is included as a supplementary material to this letter.

\section{Results \label{sec:results}}
Most studies have been focusing on Bitcoin or, at most on a few cryptocurrencies. This fact generates an overrepresentation of the big players in the literature. The analysis of eighty-four cryptocurrencies allows depicting a more comprehensive landscape of this novel and rapidly evolving market. 

Our empirical investigation is divided into two parts. The first one, computes the generalized Hurst exponent for $q=\{1,2\}$ and refines results by using a multi-scaling procedure with the computation of the curves of $qH(q)$ as a function of $q$, following the procedure developed by Di Matteo and coworkers. The second one provides more robust results by computing the generalized Hurst exponent and the multifractal spectrum of the different cryptocurrencies using the MF-DFA framework. 

The descriptive statistics of the logarithm of the average daily volume of the period, and the estimated Hurst exponents are displayed in Table \ref{tab:descriptive}.

\begin{table}
\caption{Descriptive statistics of the Log volume and generalized Hurst exponent of the sample for $q=1$ and $q=2$.}
    \centering
    \begin{tabular}{lrrr}
\toprule 
 &  & \multicolumn{2}{c}{Gen. Hurst exponent}  \\
         &Log vol.  &  $q=1$ & $q=2$ \\ 
               \midrule 
        Obs. & 84 & 84 & 84 \\ 
        Mean & 5.5210 & 0.5150 & 0.4471 \\
        Median & 6.2597 & 0.5211 & 0.4812 \\
        Min & $-0.5299$ & 0.3564 & 0.2513 \\
        Max & 8.7063 & 0.7333 & 0.5692 \\
        Std. Dev. & 1.8227 & 0.0622 & 0.0780 \\
        Skewness & $-0.9143$ & 0.0305 & $-0.8984$ \\
        Kurtosis & 3.3324 & 4.8306 & 2.5925 \\
        Jarque-Bera & 12.0904 & 11.7414 & 11.8818 \\ \bottomrule
    \end{tabular}
    \label{tab:descriptive}
\end{table}

Results regarding the estimated Hurst exponents for $q=1$ uncover an uneven behavior of cryptocurrencies. $H(1)$ describes the scaling behavior of the absolute values of the increments of a time series. 
We find that $0.5<H(1)<0.6$ for most of the largest cryptocurrencies according to traded volume. Hence, behavior is congruent with a standard Brownian motion or with a somewhat persistent stochastic process. This is in line with previous findings (referred only to Bitcoin) by \cite{Urquhart2016}, \cite{Bariviera2017}, \cite{PHILLIP201995}, and \cite{ASLAN2019}, among others.

\begin{figure}[h!p]
    \centering
        \includegraphics[scale=0.5]{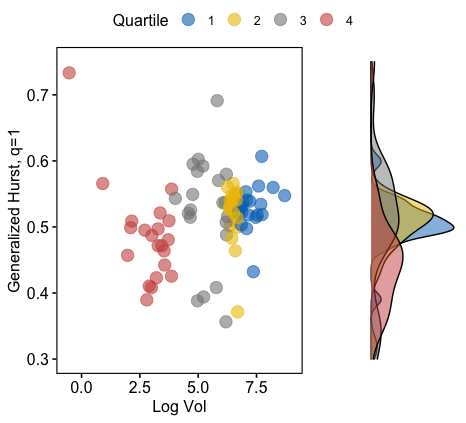}
    \caption{Scatter plot of the generalized Hurst exponent for $q=1$ and log volume of cryptocurrencies. On the right there is a density plot of the Hurst exponent, classified by volume quartile.}
   \label{fig:volGenHurstq1}
\end{figure}

From Figure \ref{fig:volGenHurstq1} it can be seen that cryptocurrencies within the third and fourth volume quartiles behave differently. Their Hurst exponents spans between $H(1)\approx 0.32$ and $H(1)\approx 0.65$. Coins in the third quartile tend to follow a persistent behavior ($H(1)>0.5$), whereas those in the fourth quartile are more likely to present an anti-persistent behavior ($H(1)<0.5$). In both cases, the time series are generally informational inefficient. A density plot is inserted on the right vertical axis of Figure \ref{fig:volGenHurstq1} in order to show the distribution of the generalized Hurst exponent in the different quartiles.

\begin{figure}[h!p]
    \centering
        \includegraphics[scale=0.5]{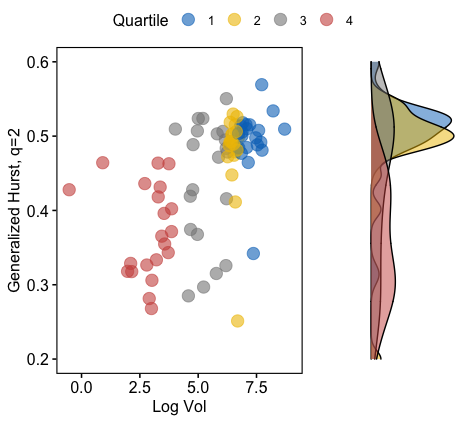}
    \caption{Scatter plot of the generalized Hurst exponent for $q=2$ and log volume of cryptocurrencies. On the right there is a density plot of the Hurst exponent, classified by volume quartile.}
   \label{fig:volGenHurstq2}
\end{figure}

The Hurst exponents for $q=2$ is connected to the autocorrelation function and connected to the power spectrum \citep{Flandrin,DiMatteo2005}. We observe, again, a noticeable behavior depending on cryptocurrency size. Cryptocurrencies in the first and second volume quantiles are roughly efficients, whereas the third and fourth quartiles exhibits an clear antipersistent behavior (see Figure \ref{fig:volGenHurstq2}).

\begin{figure}[!htbp]
    \centering
        \includegraphics[scale=0.75]{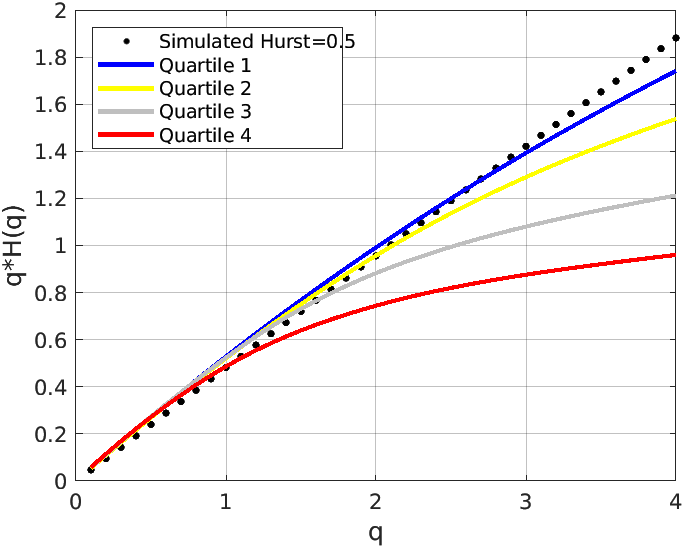}
    \caption{The function $qH(q)$ vs. $q$, averaged by quartiles, and a simulated brownian motion process with Hurst=0.5.}
   \label{fig:multifractality}
\end{figure}

As mentioned in Section \ref{sec:intro}, we generalize the analysis of the Hurst exponent for different values of $0<q<4$. Figure \ref{fig:multifractality} displays the planar representation of $q \times qH(q)$. The results of the different coins are grouped into quartiles according to volume. As a benchmark model, we include also the results arising from a simulated time series of the same length and $H=0.5$\footnote{Simulation was performed using Matlab function \texttt{wfbm}.}  If the stochastic process under consideration is monofractal (i.e. the simulated time series), $qH(q)$ as a function of $q$ is a straight line, and its slope depends on the $H$. However, the presence of nonlinearities in this function is a signature of multifractal processes. Thus, we provide compelling evidence against (fractional) Brownian, (fractional) L\`evy, and other additive, monofractal processes.

 This analysis reinforces what was shown previously regarding the heterogeneous behavior of cryptocurrencies according to their volume size. Figure \ref{fig:multifractality} clearly reveals that coins in the first quartile follow roughly unifractal processes, being a fractional brownian motion a suitable model for describing their behavior. On contrary, cryptoassets within the other cuartiles (specially those in the third and fourth) exhibit strong multifractality. In such cases,  Brownian or L\`evy models (included their fractional varieties), are deemed inadequate for capturing their complex dynamics.

\subsection{Source of multifractality: MF-DFA on original and shuffled time series}

In the previous subsection we detected multifractality in all the time series. Its presence is particularly relevant in coins and tokens belonging to the third and fourth quartiles.

In order check for the robustness of our results, and following the advise of \cite{Kukacka2020}, we compute two multifractality measures ($\Delta H, \Delta \alpha$) using MF-DFA method by \cite{Kantelhardt2002}. 
We also conduct our empirical analysis on 1000 independent shuffled realizations of each time series, and construct the 95\% confidence interval of the multifractal measures. 

\begin{figure}[!htbp]
    \centering
        \includegraphics[scale=0.7]{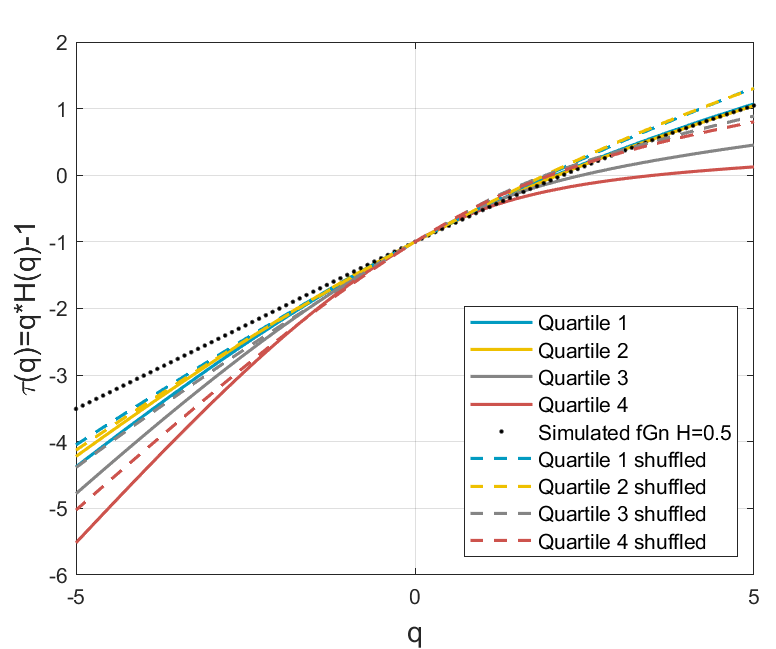}
    \caption{The function $qH(q)-1$ vs. $q$ on original and shuffled time series, averaged by quartiles. The function is also computed for a simulated brownian motion process with Hurst=0.5.}
   \label{fig:multifractality2}
\end{figure}

Figure \ref{fig:multifractality2} shows similar results to Figure \ref{fig:multifractality}, indicating that multifractality (albeit at different degrees) is present in all time series. 

Table \ref{tab:shuffling} presents the results of the two selected multifractality measures ($\Delta H,\Delta \alpha$) using the original and shuffled realizations of the time series.  Results are presented by quartiles. We observe that, even multifractality is reduced by the shuffling procedure, it does not vanish. This result indicates that the source of multifractality is (not only) a different long range correlation structure for small and large fluctuations.

We conduct also a detailed analysis of each coin and token of our sample, whose results can be found in the supplementary material to this paper. According to our results, $\Delta H$ is outside this confidence interval in 29 out of the 84 cryptocurrencies, representing 34\% of our sample. This means that in almost two-thirds of the time series multifractality is due to the correlation structure, which is destroyed by the shuffling process. Rejection rates varies according cryptocurrency size. In the first and third quartiles 28\% of the coins exhibit multifractality. In the second quartile, 24\% show results compatible with multifractal dynamics. However, in the fourth quartile, multifractality affects 57\% of the time series, meaning that an additional source of multifractality could be present. Similar results are found using $\Delta \alpha$ as multifractality measure.

\begin{table}[!htbp]
  \centering
  \caption{Multifractal measures computed on the original and shuffled time series}
    \begin{tabular}{lrrrr}
    \toprule
          & \multicolumn{2}{c}{MF-DFA} & \multicolumn{2}{c}{Multifractal spectrum} \\
          & \multicolumn{1}{l}{Original} & \multicolumn{1}{l}{Shuffled} & \multicolumn{1}{l}{Original} & \multicolumn{1}{l}{Shuffled} \\
          & \multicolumn{1}{l}{$\Delta H$} & \multicolumn{1}{l}{$\Delta H_{shuffled}$} & \multicolumn{1}{l}{$\Delta \alpha$} & \multicolumn{1}{l}{$\Delta \alpha_{shuffled}$} \\
    \midrule
    Quartile 1 & 0.2658 & 0.2016 & 0.4640 & 0.3631 \\
    Quartile 2 & 0.2628 & 0.2215 & 0.4145 & 0.3999 \\
    Quartile 3 & 0.4636 & 0.3368 & 0.7344 & 0.5576 \\
    Quartile 4 & 0.6825 & 0.4708 & 1.0162 & 0.7374 \\
    \bottomrule
    \end{tabular}%
  \label{tab:shuffling}%
\end{table}%

Subsequently, we explore the role of kurtosis in the multifractal profile of the time series. Kurtosis is a common descriptive statistic to signal the peakedness of the probability density function and the presence of fat tails. It was previously reported \citep{Corbet2019,Aslanidis2019} that cryptocurrencies exhibit stronger fluctuations than traditional assets. This could be specially true for the smallest and most illiquid assets, which are prone to sudden jumps. 

\begin{figure}[htbp!]
    \centering
        \includegraphics[scale=0.7]{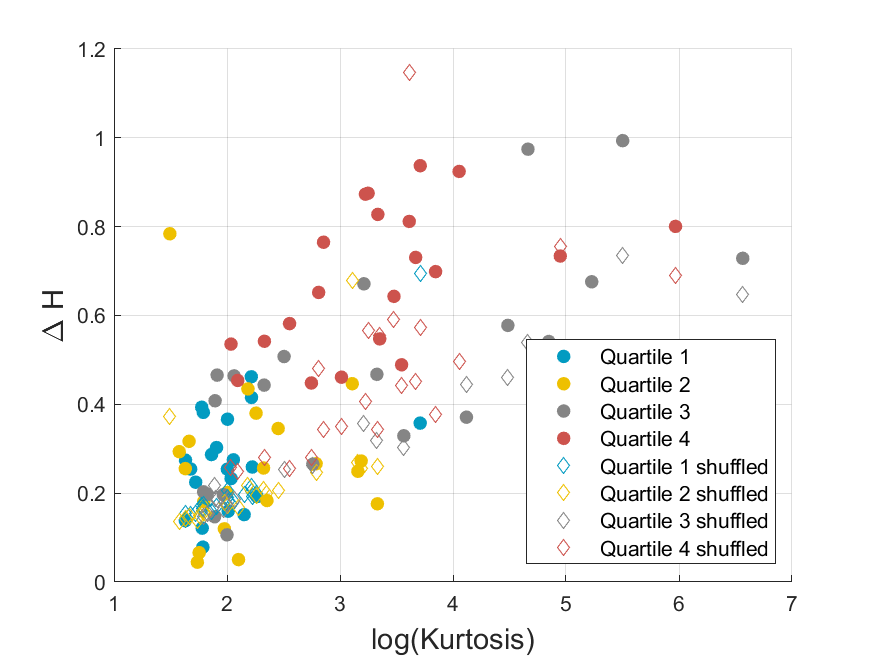}
    \caption{Relationship between kurtosis and $\Delta H$ as a measure of the level of multifractality. Each point represents a cryptocurrency using the original and the shuffled time series.}
   \label{fig:KurtosisH}
\end{figure}

\begin{figure}[htbp!]
    \centering
        \includegraphics[scale=0.7]{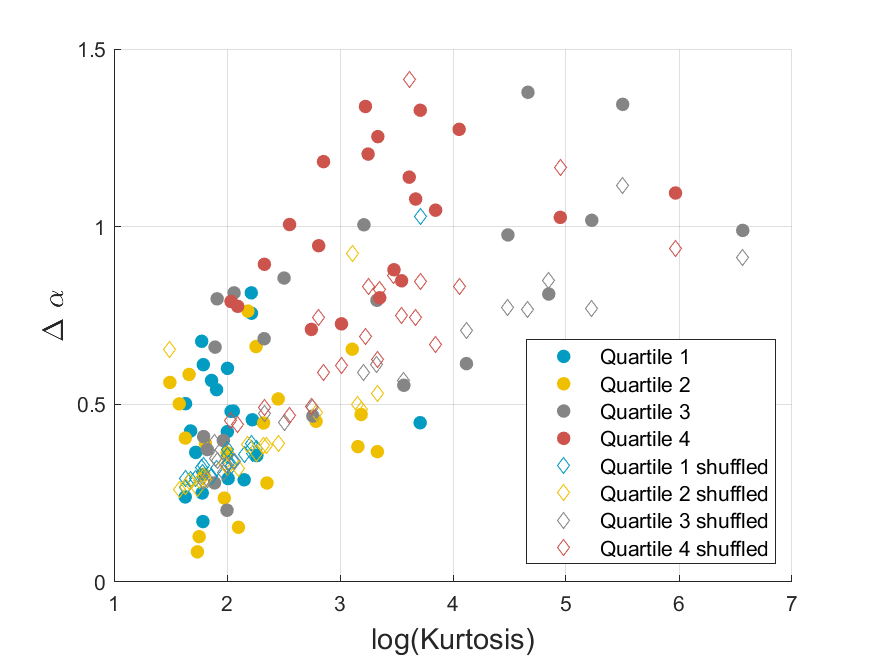}
    \caption{Relationship between kurtosis and $\Delta \alpha$ as a measure of the level of multifractality. Each point represents a cryptocurrency using the original and the shuffled time series.}
   \label{fig:KurtosisAlpha}
\end{figure}

Figure \ref{fig:KurtosisH} and \ref{fig:KurtosisAlpha} show an association between the measures of multifractality ($\Delta H, \Delta \alpha$) and the estimated kurtosis of the time series. This association is (to a great extent) destroyed by the shuffling procedure in coins and tokens belonging to the first and second quartiles. Thus,  kurtosis seems not to influence significantly in the level of multifractality of these coins. This result is consistent with our previous finding that only 28\% (23\%) of the cryptocurrencies in the first (second) quartile lie in the critical region of multifractality. However, extreme events (proxied by greater kurtosis) in the third and fourth quartiles seem to translate into stronger multifractality. As a caveat, we are not saying that these time series are less informational efficient, but that smaller coins and tokens present more frequent and larger jumps, which in turn induces more multifractality. 

\section{Conclusions \label{sec:conclusions}}

This letter sheds light on the multifractal behavior of the cryptocurrency market in a broad way. 

We expand previous research computing the generalized Hurst exponent and the multifractal spectrum of eighty-four cryptocurrencies time series using two alternative methods. 

According to our results, cryptocurrencies have a different long memory endowment, according to their size, proxied by traded volume. More importantly, we detected the presence of multiflactality in several time series. Largest cryptocurrencies (those in the first quartile of volume) seem to follow monofractal processes, consistent with a fractional Brownian motion. On contrary, other cryptocurrencies exhibit strong multifractality. This result poses some restrictions on the suitable stochastic models for such coins and tokens. 

Regarding the source of multifractality, our results are mixed. According to our results, shuffling the time series produces a reduction in the level of multifractality, but does not offset all of it. This result is in agreement, regarding bitcoin, with \cite{Kukacka2020}. Consequently, there is another source of multifractality. We find an association of kurtosis with greater levels of multifractality. This association is stronger for the smallest coins and tokens of our sample.  

Consequently, our results support the idea that cryptocurrencies differ not only among them in their long range dependence, but also in the stochastic processes that govern their dynamical behavior.

Our findings can be of interest for academics and practitioners alike. From the academic point of view, means that one model does not fit all. It is necessary to study on a case-by-case basis, in order to select the most appropriate model to describe return dynamics. From the practitioners point of view, means that there could be some arbitrage opportunities, depending on each cryptocurrency.

\bibliographystyle{apalike}
\bibliography{longmemorybib}

\appendix
\section*{Supplementary material}
\newgeometry{top=1in,bottom=1in,right=0.5in,left=0.5in}
\footnotesize
\begin{longtable}{lrrrrrrrrr}
  \caption{Descriptive statistics of the logarithmic returns of the cryptocurrencies in the sample.}   \\
   \hline
Crypto & Obs. & Mean & Median  & Min & Max & Std. Dev. & Skewness & Kurtosis & Jarque-Bera  \\
\hline \\
\endfirsthead
\multicolumn{4}{c} {\tablename\ \thetable\ -- \textit{Continued from previous page}} \\
\hline     
Crypto & Obs. & Mean & Median & Min  & Max & Std. Dev. & Skewness & Kurtosis & Jarque-Bera \\
\hline
\endhead
\hline \multicolumn{4}{r}{\textit{Continued on next page}} \\
\endfoot
\hline
\endlastfoot
ACHN   & 789 & -0.6465 & -0.4455 & -31.2524  & 40.3689  & 7.3853   & 0.0643   & 6.7701   & 467.8120      \\
ACOIN  & 789  & 0.0222  & -0.0924 & -642.0136 & 786.4650 & 36.6010   & 5.7431   & 391.1575 & 4957488.9982  \\
ADA    & 789  & -0.3820 & 0.0000  & -29.1399  & 27.4731  & 5.8657  & -0.1756  & 6.1123   & 322.4990      \\
AION   & 789  & -0.5223 & -0.4504 & -32.3820  & 30.2058  & 7.3928   & 0.0451   & 5.1096   & 146.5718      \\
ARG    & 789  & 0.1121  & -0.0877 & -100.7488 & 141.5843 & 12.7185    & 1.7287   & 39.2419  & 43573.4217    \\
ARI    & 789  & -0.3161 & -0.1309 & -43.0557  & 82.9262  & 9.6098   & 1.4270   & 17.3601  & 7046.9759     \\
ARN    & 789  & -0.4634 & -0.3547 & -32.7932  & 53.0574  & 7.7215  & 0.3605   & 8.1794   & 899.0122      \\
BAT    & 789  & -0.1312 & -0.0569 & -36.9243  & 28.1515  & 6.4885   & -0.2420  & 5.6834   & 244.4252      \\
BCD    & 789  & -0.5210 & -0.5324 & -219.5634 & 225.5275 & 34.7946  & 0.1833   & 22.3872  & 12360.9851    \\
BCH    & 789  & -0.2554 & -0.3140 & -30.6758  & 41.9398  & 6.5121  & 0.4938   & 9.1610   & 1279.9402     \\
BET    & 789  & -0.2057 & -0.0969 & -151.7714 & 100.6797 & 15.5515  & -0.8957  & 28.5438  & 21555.9288    \\
BNB    & 789  & -0.0119 & -0.0153 & -36.4052  & 23.4764  & 5.3866  & -0.4235  & 7.6622   & 738.1623      \\
BOST   & 789  & -0.1515 & 0.0245  & -80.7112  & 94.6135  & 9.8526  & 0.9534   & 32.3838  & 28504.0433    \\
BTB    & 789 & -0.4194 & -0.5946 & -112.7231 & 82.0514  & 13.8699 & -0.1622  & 12.8555  & 3196.6665     \\
BTC    & 789  & -0.0808 & 0.0608  & -18.9167  & 16.7222  & 3.9346  & -0.2803  & 5.9933   & 304.8868      \\
BTCD   & 789  & -0.3552 & -0.1088 & -130.8021 & 74.8189  & 8.4887  & -2.8383  & 88.7494  & 242788.1233   \\
BTM    & 789  & -0.2498 & -0.1563 & -37.8147  & 69.1948  & 7.1966  & 0.8310   & 16.2541  & 5865.9775     \\
BTMK   & 789  & -0.6063 & -0.0613 & -194.7520 & 183.8111 & 15.9052  & -1.3350  & 105.9850 & 348903.4458   \\
CACH   & 789  & -0.3160 & -0.0743 & -78.0010  & 75.5404  & 7.8412  & -0.1794  & 28.0759  & 20676.0963    \\
CANN   & 789  & -0.4099 & -0.3122 & -152.6686 & 163.8534 & 14.4065  & 1.2916   & 61.5005  & 112727.8183   \\
CAP    & 789  & -0.2484 & 0.0626  & -426.2745 & 408.5881 & 52.2362  & 0.3860   & 37.0638  & 38165.8014    \\
CASH   & 789  & -0.5327 & 0.0390  & -317.4608 & 127.6671 & 17.6535   & -6.9373  & 141.1032 & 633337.1767   \\
CBX    & 789 & -0.5713 & -0.1947 & -173.0365 & 284.3819 & 20.9427  & 2.0769   & 57.6469  & 98741.4534    \\
CCN    & 789  & -0.3522 & 0.0318  & -35.7165  & 32.9575  & 6.4058  & -0.4773  & 8.1235   & 892.9247      \\
CLAM   & 789 & -0.3202 & -0.0927 & -91.0199  & 49.0385  & 7.7295 & -2.6068  & 35.3349  & 35265.8323    \\
CVC    & 789  & -0.4552 & -0.0531 & -35.6778  & 32.2799  & 6.4506  & -0.3332  & 6.2040   & 352.0709      \\
DASH   & 789  & -0.3241 & -0.2916 & -22.8908  & 39.8701  & 5.5138   & 0.5476   & 9.5572   & 1452.9624     \\
DGC    & 789  & -0.3833 & 0.0000  & -37.6144  & 36.3121  & 8.4641 & -0.3107  & 7.6549   & 725.0299      \\
ELF    & 789  & -0.3752 & -0.0621 & -29.6456  & 36.0310  & 7.2397   & 0.0800   & 5.9060   & 278.4677      \\
EMC2   & 789  & -0.3265 & -0.3665 & -51.1030  & 51.1508  & 10.2620   & 0.1591   & 10.2620  & 1737.0282     \\
ENJ    & 789  & -0.1772 & -0.2164 & -42.9976  & 77.9986  & 8.2509    & 2.0260   & 23.5325  & 14399.3072    \\
ENRG   & 789  & -0.4763 & -0.1491 & -70.7377  & 30.6517  & 6.1280   & -2.0585  & 27.8445  & 20849.3470    \\
EOS    & 789  & -0.1214 & -0.0702 & -28.7411  & 35.0657  & 6.5429   & 0.3024   & 7.4113   & 651.7630      \\
ETC    & 789  & -0.1801 & -0.0831 & -36.2296  & 30.7827  & 5.7578    & -0.2930  & 7.4205   & 653.7058      \\
ETH    & 789  & -0.1878 & -0.1019 & -22.2767  & 17.9934  & 5.0447    & -0.3609  & 5.1191   & 164.7538      \\
GEO    & 789  & -0.5341 & -0.3805 & -331.3922 & 339.6577 & 20.8935        & 1.9021   & 186.3900 & 1106124.2826  \\
GNT    & 789  & -0.3697 & -0.1605 & -36.3795  & 52.2172  & 6.5724         & 0.2528   & 10.5234  & 1869.1748     \\
GTO    & 789  & -0.5140 & -0.4676 & -43.2038  & 84.4657  & 9.1536         & 1.1219   & 15.7960  & 5548.3959     \\
ICX    & 789  & -0.3776 & -0.1574 & -60.7342  & 40.8273  & 8.0278         & -0.2359  & 9.6235   & 1449.5530     \\
KCS    & 789  & -0.2935 & -0.3150 & -28.3733  & 29.8662  & 6.3389         & 0.0947   & 6.6498   & 439.1097      \\
KNC    & 789 & -0.2219 & 0.0000  & -55.2516  & 58.1159  & 11.8545    & -0.0603  & 7.2242   & 587.0904      \\
LIMX   & 789  & -0.0764 & 0.0415  & -92.1752  & 109.5124 & 10.1987   & 0.9206   & 40.8651  & 47246.4305    \\
LRC    & 789  & -0.4255 & -0.0827 & -30.5002  & 41.9089  & 7.2161    & 0.1971   & 7.1533   & 572.1954      \\
LSK    & 789  & -0.3935 & -0.4529 & -27.3164  & 26.3976  & 6.0197    & 0.0797   & 5.9320   & 283.4470      \\
LTC    & 789  & -0.1903 & -0.3236 & -23.2089  & 28.7415  & 5.2683    & 0.3302   & 6.4429   & 404.0318      \\
MANA   & 789  & -0.1692 & 0.0743  & -61.6632  & 79.6635  & 7.6853    & 0.8800   & 24.2182  & 14902.5535    \\
MBL    & 789  & 0.1879  & 0.0000  & -113.0223 & 318.6700 & 15.3015   & 10.9820  & 245.0695 & 1942257.4954  \\
MCO    & 789  & -0.1420 & -0.1146 & -35.2500  & 46.1690  & 6.3769    & 0.1502   & 9.2438   & 1284.6039     \\
MIOTA  & 789  & -0.3627 & -0.2534 & -30.8061  & 22.5011  & 5.7812   & -0.2587  & 5.3545   & 191.0468      \\
MONA   & 789  & -0.1915 & -0.4290 & -39.4149  & 67.6841  & 7.6719    & 2.5334   & 24.8191  & 16494.8923    \\
MOON   & 789  & -0.5358 & 0.0000  & -48.9576  & 24.0866  & 6.6536  & -1.5572  & 12.2443  & 3128.2807     \\
NANO   & 789  & -0.4632 & -0.4115 & -41.6924  & 34.0628  & 7.4696    & -0.1879  & 7.4662   & 660.4142      \\
NEO    & 789  & -0.2676 & -0.1794 & -29.1631  & 25.2006  & 6.1411  & -0.0023  & 5.6009   & 222.3978      \\
OMG    & 789  & -0.3848 & -0.2395 & -26.3033  & 23.9605  & 6.0345    & -0.2443  & 5.1070   & 153.8039      \\
POLY   & 789  & 0.6466  & -0.0927 & -74.0112  & 860.4205 & 31.4587    & 25.9158  & 709.0048 & 16474626.7901 \\
POWR   & 789  & -0.3455 & 0.0000  & -35.9366  & 33.6455  & 6.5772    & 0.1750   & 7.2152   & 588.1475      \\
PRC    & 789  & -0.3727 & -0.0394 & -126.7511 & 87.2883  & 11.9586   & -1.6035  & 34.6372  & 33243.0541    \\
QRK    & 789  & -0.2496 & 0.0000  & -47.9456  & 62.0739  & 7.5338    & 0.1203   & 15.5921  & 5214.6090     \\
QTL    & 789  & -0.2532 & -0.1325 & -30.3338  & 45.2547  & 6.6903    & 0.2904   & 10.2879  & 1757.1806     \\
QTUM   & 789  & -0.4422 & -0.2782 & -37.5255  & 39.7683  & 6.4722    & 0.0967   & 7.8155   & 763.5804     \\
REP    & 789  & -0.2296 & -0.0900 & -31.6676  & 51.2594  & 6.3314    & 0.8388   & 11.6203  & 2535.4303    \\
RIC    & 789  & -0.4812 & 0.0375  & -149.0996 & 35.4242  & 8.5759    & -7.8577  & 127.3983 & 516857.8624  \\
SNT    & 789  & -0.4410 & -0.2741 & -28.9278  & 28.1204  & 5.5881    & -0.2170  & 5.7765   & 259.6161  \\
SRN    & 789  & -0.6145 & -0.4791 & -58.3293  & 48.7632  & 7.9451   & -0.2207  & 10.2111  & 1715.8862  \\
STEEM  & 789  & -0.4478 & -0.4249 & -35.5088  & 37.8818  & 6.4200  & 0.1345   & 7.3960   & 637.6729      \\
STORM  & 789  & -0.5849 & -0.6433 & -35.9387  & 99.2795  & 8.5338  & 1.8081   & 27.9611  & 20912.8677    \\
SWFTC  & 789  & -0.4765 & -0.3775 & -37.0445  & 47.7890  & 7.8494  & 0.2390   & 8.8924   & 1148.9625     \\
SXC    & 789  & -0.6798 & -0.2708 & -85.5243  & 120.6038 & 12.7925  & 0.5567   & 20.3478  & 9934.3193     \\
TRX    & 789  & -0.2870 & -0.1965 & -34.7714  & 36.9231  & 6.5874   & -0.0859  & 6.7353   & 459.6622      \\
VET    & 789 & -0.2584 & -0.0532 & -309.1562 & 233.9956 & 26.9495   & -1.1687  & 40.8282  & 47222.8119    \\
WAVES  & 789  & -0.2865 & -0.2147 & -26.6230  & 38.3516  & 6.0615  & 0.3906   & 8.5999   & 1050.9990     \\
WTC    & 789  & -0.4396 & -0.4270 & -29.5165  & 43.5201  & 7.3645  & 0.2089   & 6.1065   & 322.9881  \\
XBS    & 789 & -0.2713 & 0.0348  & -89.3732  & 66.6890  & 9.5751         & -0.7232  & 25.7608  & 17099.8341  \\
XEM    & 789  & -0.4316 & -0.1763 & -29.8042  & 24.0098  & 5.7501    & -0.0069  & 5.9686   & 289.7271      \\
XLM    & 789 & -0.3096 & -0.2718 & -32.9406  & 26.8116  & 5.6059 & -0.0605  & 6.0157   & 299.4597      \\
XMR    & 789  & -0.2213 & 0.0153  & -27.6457  & 18.5045  & 5.3125   & -0.3864  & 5.2822   & 190.8571  \\
XMY    & 789  & -0.4934 & -0.3226 & -48.9548  & 28.0314  & 7.7976    & -0.4281  & 6.0080   & 321.5598   \\
XPY    & 789  & -0.2489 & 0.0656  & -68.3288  & 66.1725  & 10.5757        & -0.2512  & 16.6274  & 6113.4152     \\
XRP    & 789  & -0.3047 & -0.2841 & -36.7056  & 31.7479  & 5.3604     & 0.0917   & 9.1769   & 1255.4367     \\
XTZ    & 789  & -0.0679 & 0.0000  & -99.9714  & 52.2926  & 7.9993     & -3.7663  & 51.7874  & 80114.7553    \\
XUC    & 789  & -0.2481 & -0.2478 & -25.2063  & 29.2463  & 5.1009     & 0.3328   & 7.8677   & 793.5280      \\
YBC    & 789  & -0.0960 & 0.0624  & -68.3073  & 35.7470  & 5.2919     & -2.2015  & 46.8009  & 63708.4791    \\
ZEC    & 789   & -0.3217 & -0.3374 & -21.1347  & 22.3873  & 5.4391    & 0.0198   & 4.8426   & 111.6624  \\
ZET    & 789  & -0.3848 & 0.0121  & -57.4514  & 87.3608  & 9.2349     & 1.0451   & 25.1554  & 16280.7154   
\end{longtable}

\section{Multifractal measures for each cryptocurency.}

{\small
\begin{longtable}{lrrrrl}
  \caption{Difference between the maximum and minimum Generalized Hurst exponent computed by means of MF-DFA method, on the original series and on 1000 shuffled resampling of each series. Average and confidence limits at 0.025 and 0.975 percentiles are reported for the shuffled series. * means that $\Delta H$ and $\Delta H_{shuffled}$ differ at 95\% confidence level.}   \\
   \hline
& \multicolumn{1}{l}{Original Series} & \multicolumn{4}{c}{Shuffled series} \\
    Crypto & \multicolumn{1}{r}{$\Delta H$} & \multicolumn{1}{r}{$\Delta H_{shuffled}$} & \multicolumn{1}{r}{CL\_0.025} & \multicolumn{1}{r}{CL\_0.975} &  \\
\hline \\
\endfirsthead
\multicolumn{4}{c} {\tablename\ \thetable\ -- \textit{Continued from previous page}} \\
\hline     
& \multicolumn{1}{l}{Original Series} & \multicolumn{4}{c}{Shuffled series} \\
    Crypto & \multicolumn{1}{r}{$\Delta H$} & \multicolumn{1}{r}{$\Delta H_{shuffled}$} & \multicolumn{1}{r}{CL\_0.025} & \multicolumn{1}{r}{CL\_0.975} &  \\
		\hline
\endhead
\hline \multicolumn{4}{r}{\textit{Continued on next page}} \\
\endfoot
\hline
\endlastfoot
  ACHN  & 0.4658 & 0.1849 & 0.0388 & 0.3642 & * \\
    ACOIN & 0.8007 & 0.6895 & 0.4157 & 0.8594 &  \\
    ADA   & 0.1062 & 0.1768 & 0.0378 & 0.3464 &  \\
    AION  & 0.2553 & 0.1425 & 0.0210 & 0.3049 &  \\
    ARG   & 0.7308 & 0.4525 & 0.1765 & 0.6852 & * \\
    ARI   & 0.7649 & 0.3424 & 0.1076 & 0.5789 & * \\
    ARN   & 0.0504 & 0.1667 & 0.0300 & 0.3385 &  \\
    BAT   & 0.0446 & 0.1383 & 0.0247 & 0.2924 &  \\
    BCD   & 0.4464 & 0.6785 & 0.3549 & 0.9696 &  \\
    BCH   & 0.4621 & 0.2140 & 0.0465 & 0.3992 & * \\
    BET   & 0.5473 & 0.5544 & 0.2586 & 0.8160 &  \\
    BNB   & 0.2330 & 0.1736 & 0.0330 & 0.3531 &  \\
    BOST  & 0.6431 & 0.5908 & 0.3168 & 0.8566 &  \\
    BTB   & 0.5818 & 0.2566 & 0.0597 & 0.4622 & * \\
    BTC   & 0.3819 & 0.1745 & 0.0298 & 0.3541 & * \\
    BTCD  & 0.5777 & 0.4611 & 0.2045 & 0.7141 &  \\
    BTM   & 0.2664 & 0.2459 & 0.0621 & 0.4397 &  \\
    BTMK  & 0.9745 & 0.5391 & 0.3082 & 0.7784 & * \\
    CACH  & 0.8276 & 0.3431 & 0.1039 & 0.5606 & * \\
    CANN  & 0.3710 & 0.4457 & 0.2051 & 0.6757 &  \\
    CAP   & 0.8118 & 1.1464 & 0.6598 & 1.6088 &  \\
    CASH  & 0.7339 & 0.7552 & 0.3911 & 1.1027 &  \\
    CBX   & 0.9244 & 0.4965 & 0.2320 & 0.7701 & * \\
    CCN   & 0.4538 & 0.2485 & 0.0685 & 0.4449 & * \\
    CLAM  & 0.3291 & 0.3024 & 0.0930 & 0.5187 &  \\
    CVC   & 0.1970 & 0.1547 & 0.0274 & 0.3122 &  \\
    DASH  & 0.3802 & 0.1949 & 0.0456 & 0.3704 & * \\
    DGC   & 0.5357 & 0.2575 & 0.0684 & 0.4600 & * \\
    ELF   & 0.3935 & 0.1716 & 0.0332 & 0.3380 & * \\
    EMC2  & 0.4433 & 0.2802 & 0.0783 & 0.4855 &  \\
    ENJ   & 0.2493 & 0.2697 & 0.0707 & 0.4831 &  \\
    ENRG  & 0.4676 & 0.3194 & 0.0896 & 0.5550 &  \\
    EOS   & 0.2540 & 0.2056 & 0.0391 & 0.3936 &  \\
    ETC   & 0.3665 & 0.1908 & 0.0337 & 0.3803 &  \\
    eth   & 0.2742 & 0.1540 & 0.0267 & 0.3256 &  \\
    GEO   & 0.6759 & 0.5253 & 0.3002 & 0.7448 &  \\
    GNT   & 0.1835 & 0.2002 & 0.0368 & 0.3869 &  \\
    GTO   & 0.2653 & 0.2621 & 0.0670 & 0.4683 &  \\
    ICX   & 0.1922 & 0.1997 & 0.0432 & 0.3854 &  \\
    KCS   & 0.4080 & 0.1888 & 0.0355 & 0.3794 & * \\
    KNC   & 0.1469 & 0.2170 & 0.0462 & 0.4192 &  \\
    LIMX  & 0.9372 & 0.5721 & 0.2860 & 0.8504 & * \\
    LRC   & 0.1967 & 0.1638 & 0.0275 & 0.3240 &  \\
    LSK   & 0.1213 & 0.1599 & 0.0288 & 0.3236 &  \\
    LTC   & 0.2869 & 0.1580 & 0.0276 & 0.3134 &  \\
    MANA  & 0.2724 & 0.2546 & 0.0692 & 0.4578 &  \\
    MBL   & 0.9937 & 0.7361 & 0.3659 & 1.1271 &  \\
    MCO   & 0.2589 & 0.1918 & 0.0406 & 0.3822 &  \\
    MIOTA & 0.2544 & 0.1518 & 0.0279 & 0.3141 &  \\
    MONA  & 0.6713 & 0.3557 & 0.1234 & 0.5675 & * \\
    MOON  & 0.5076 & 0.2525 & 0.0687 & 0.4437 & * \\
    NANO  & 0.1593 & 0.1813 & 0.0309 & 0.3749 &  \\
    NEO   & 0.2247 & 0.1547 & 0.0253 & 0.3184 &  \\
    OMG   & 0.1380 & 0.1405 & 0.0231 & 0.3044 &  \\
    POLY  & 0.7288 & 0.6480 & 0.2831 & 1.4736 &  \\
    POWR  & 0.1198 & 0.1791 & 0.0378 & 0.3583 &  \\
    PRC   & 0.4891 & 0.4428 & 0.1793 & 0.7326 &  \\
    QRK   & 0.4480 & 0.2797 & 0.0693 & 0.4889 &  \\
    QTL   & 0.5420 & 0.2802 & 0.0758 & 0.4820 & * \\
    QTUM  & 0.2751 & 0.1811 & 0.0306 & 0.3520 &  \\
    REP   & 0.3456 & 0.2050 & 0.0450 & 0.3998 &  \\
    RIC   & 0.5415 & 0.5244 & 0.2293 & 0.8086 &  \\
    SNT   & 0.0661 & 0.1507 & 0.0264 & 0.3115 &  \\
    SRN   & 0.2566 & 0.2077 & 0.0473 & 0.3930 &  \\
    STEEM & 0.2016 & 0.1733 & 0.0315 & 0.3529 &  \\
    STORM & 0.1761 & 0.2611 & 0.0663 & 0.4711 &  \\
    SWFTC & 0.4346 & 0.2163 & 0.0389 & 0.4068 & * \\
    SXC   & 0.4611 & 0.3507 & 0.1018 & 0.5700 &  \\
    TRX   & 0.3027 & 0.1699 & 0.0277 & 0.3390 &  \\
    VET   & 0.3577 & 0.6949 & 0.3828 & 0.9865 & * \\
    WAVES & 0.1517 & 0.1964 & 0.0355 & 0.3722 &  \\
    WTC   & 0.1938 & 0.1532 & 0.0247 & 0.3034 &  \\
    XBS   & 0.8752 & 0.5660 & 0.2886 & 0.8291 & * \\
    XEM   & 0.0785 & 0.1648 & 0.0274 & 0.3255 &  \\
    XLM   & 0.1812 & 0.1561 & 0.0245 & 0.3283 &  \\
    XMR   & 0.3167 & 0.1485 & 0.0296 & 0.3149 & * \\
    XMY   & 0.2030 & 0.1478 & 0.0248 & 0.3003 &  \\
    XPY   & 0.6519 & 0.4796 & 0.2259 & 0.7469 &  \\
    XRP   & 0.4156 & 0.2043 & 0.0424 & 0.3864 & * \\
    XTZ   & 0.7841 & 0.3719 & 0.1418 & 0.5943 & * \\
    XUC   & 0.4641 & 0.1861 & 0.0381 & 0.3596 & * \\
    YBC   & 0.6988 & 0.3765 & 0.1406 & 0.5887 & * \\
    ZEC   & 0.2934 & 0.1370 & 0.0255 & 0.3045 &  \\
    ZET   & 0.8732 & 0.4054 & 0.1568 & 0.6558 & * \\
		\end{longtable}
}

{\small
\begin{longtable}{lrrrrl}
  \caption{Difference between the maximum and minimum value in the multifractal spectrum computed by means of MF-DFA method, on the original series and on 1000 shuffled resampling of each series. Average and confidence limits at 0.025 and 0.975 percentiles are reported for the shuffled series. * means that $\Delta \alpha$ and $\Delta \alpha_{shuffled}$ differ at 95\% confidence level.}   \\
   \hline
& \multicolumn{1}{l}{Original Series} & \multicolumn{4}{c}{Shuffled series} \\
    Crypto & \multicolumn{1}{r}{$\Delta \alpha$} & \multicolumn{1}{r}{$\Delta \alpha_{shuffled}$} & \multicolumn{1}{r}{CL\_0.025} & \multicolumn{1}{r}{CL\_0.975} &  \\
\hline \\
\endfirsthead
\multicolumn{4}{c} {\tablename\ \thetable\ -- \textit{Continued from previous page}} \\
\hline     
& \multicolumn{1}{l}{Original Series} & \multicolumn{4}{c}{Shuffled series} \\
    Crypto & \multicolumn{1}{r}{$\Delta \alpha$} & \multicolumn{1}{r}{$\Delta \alpha_{shuffled}$} & \multicolumn{1}{r}{CL\_0.025} & \multicolumn{1}{r}{CL\_0.975} &  \\
		\hline
\endhead
\hline \multicolumn{4}{r}{\textit{Continued on next page}} \\
\endfoot
\hline
\endlastfoot
    ACHN  & 0.7969 & 0.3418 & 0.0887 & 0.6469 & * \\
    ACOIN & 1.0948 & 0.9389 & 0.5907 & 1.1800 &  \\
    ADA   & 0.2020 & 0.3297 & 0.0836 & 0.6304 &  \\
    AION  & 0.4050 & 0.2689 & 0.0514 & 0.5503 &  \\
    ARG   & 1.0777 & 0.7446 & 0.3507 & 1.1190 &  \\
    ARI   & 1.1830 & 0.5892 & 0.2177 & 0.9763 & * \\
    ARN   & 0.1539 & 0.3182 & 0.0717 & 0.6219 &  \\
    BAT   & 0.0849 & 0.2634 & 0.0587 & 0.5508 &  \\
    BCD   & 0.6551 & 0.9240 & 0.4756 & 1.3677 &  \\
    BCH   & 0.8137 & 0.3886 & 0.1130 & 0.7037 & * \\
    BET   & 0.7998 & 0.8237 & 0.4123 & 1.2120 &  \\
    BNB   & 0.4802 & 0.3287 & 0.0746 & 0.6405 &  \\
    BOST  & 0.8788 & 0.8638 & 0.4600 & 1.2608 &  \\
    BTB   & 1.0060 & 0.4689 & 0.1436 & 0.8148 & * \\
    BTC   & 0.6117 & 0.3267 & 0.0747 & 0.6370 &  \\
    BTCD  & 0.9766 & 0.7720 & 0.4001 & 1.1553 &  \\
    BTM   & 0.4522 & 0.4782 & 0.1472 & 0.8102 &  \\
    BTMK  & 1.3781 & 0.7681 & 0.4527 & 1.0919 & * \\
    CACH  & 1.2534 & 0.6259 & 0.2590 & 0.9828 & * \\
    CANN  & 0.6145 & 0.7092 & 0.3694 & 1.0539 &  \\
    CAP   & 1.1393 & 1.4133 & 0.8233 & 2.0186 &  \\
    CASH  & 1.0262 & 1.1660 & 0.6644 & 1.6672 &  \\
    CBX   & 1.2739 & 0.8320 & 0.4582 & 1.2537 & * \\
    CCN   & 0.7758 & 0.4429 & 0.1445 & 0.7788 &  \\
    CLAM  & 0.5537 & 0.5661 & 0.2399 & 0.9093 &  \\
    CVC   & 0.3727 & 0.2894 & 0.0605 & 0.5474 &  \\
    DASH  & 0.6627 & 0.3613 & 0.0979 & 0.6634 &  \\
    DGC   & 0.7895 & 0.4555 & 0.1462 & 0.7890 & * \\
    ELF   & 0.6774 & 0.3210 & 0.0791 & 0.6197 & * \\
    EMC2  & 0.6846 & 0.4744 & 0.1537 & 0.8179 &  \\
    ENJ   & 0.3810 & 0.4993 & 0.1820 & 0.8634 &  \\
    ENRG  & 0.7929 & 0.6132 & 0.2362 & 0.9990 &  \\
    EOS   & 0.4229 & 0.3739 & 0.0929 & 0.6960 &  \\
    ETC   & 0.6013 & 0.3572 & 0.0730 & 0.6905 &  \\
    ETH   & 0.5019 & 0.2921 & 0.0638 & 0.5947 &  \\
    GEO   & 1.0179 & 0.7685 & 0.4879 & 1.0676 &  \\
    GNT   & 0.2786 & 0.3828 & 0.0913 & 0.7199 &  \\
    GTO   & 0.4671 & 0.4907 & 0.1584 & 0.8531 &  \\
    ICX   & 0.3558 & 0.3745 & 0.1008 & 0.6926 &  \\
    KCS   & 0.6611 & 0.3469 & 0.0834 & 0.6633 &  \\
    KNC   & 0.2789 & 0.3925 & 0.0962 & 0.7341 &  \\
    LIMX  & 1.3277 & 0.8452 & 0.4525 & 1.2614 & * \\
    LRC   & 0.3982 & 0.3080 & 0.0679 & 0.5949 &  \\
    LSK   & 0.2505 & 0.2973 & 0.0617 & 0.5697 &  \\
    LTC   & 0.5672 & 0.2962 & 0.0647 & 0.5797 &  \\
    MANA  & 0.4714 & 0.4863 & 0.1767 & 0.8241 &  \\
    MBL   & 1.3442 & 1.1166 & 0.6022 & 1.6547 &  \\
    MCO   & 0.4566 & 0.3610 & 0.0918 & 0.6705 &  \\
    MIOTA & 0.4251 & 0.2882 & 0.0623 & 0.5714 &  \\
    MONA  & 1.0053 & 0.5892 & 0.2350 & 0.9335 & * \\
    MOON  & 0.8555 & 0.4494 & 0.1535 & 0.7891 & * \\
    NANO  & 0.2912 & 0.3331 & 0.0738 & 0.6733 &  \\
    NEO   & 0.3648 & 0.2908 & 0.0660 & 0.6015 &  \\
    OMG   & 0.2394 & 0.2669 & 0.0532 & 0.5616 &  \\
    POLY  & 0.9894 & 0.9140 & 0.4314 & 1.9224 &  \\
    POWR  & 0.2361 & 0.3325 & 0.0836 & 0.6410 &  \\
    PRC   & 0.8478 & 0.7494 & 0.3607 & 1.1958 &  \\
    QRK   & 0.7108 & 0.4926 & 0.1675 & 0.8468 &  \\
    QTL   & 0.8941 & 0.4897 & 0.1663 & 0.8228 & * \\
    QTUM  & 0.4810 & 0.3377 & 0.0695 & 0.6345 &  \\
    REP   & 0.5153 & 0.3887 & 0.1088 & 0.7424 &  \\
    RIC   & 0.8104 & 0.8473 & 0.4180 & 1.2594 &  \\
    SNT   & 0.1277 & 0.2849 & 0.0585 & 0.5751 &  \\
    SRN   & 0.4476 & 0.3839 & 0.0950 & 0.7032 &  \\
    STEEM & 0.3634 & 0.3236 & 0.0693 & 0.6429 &  \\
    STORM & 0.3669 & 0.5313 & 0.1864 & 0.9052 &  \\
    SWFTC & 0.7625 & 0.3870 & 0.0985 & 0.7094 & * \\
    SXC   & 0.7265 & 0.6082 & 0.2143 & 0.9679 &  \\
    TRX   & 0.5413 & 0.3188 & 0.0672 & 0.6215 &  \\
    VET   & 0.4482 & 1.0281 & 0.5941 & 1.4569 & * \\
    WAVES & 0.2877 & 0.3586 & 0.0910 & 0.6614 &  \\
    WTC   & 0.3899 & 0.2898 & 0.0631 & 0.5498 &  \\
    XBS   & 1.2043 & 0.8312 & 0.4218 & 1.2404 &  \\
    XEM   & 0.1701 & 0.3095 & 0.0641 & 0.6034 &  \\
    XLM   & 0.3031 & 0.2970 & 0.0615 & 0.5986 &  \\
    XMR   & 0.5842 & 0.2818 & 0.0687 & 0.5818 & * \\
    XMY   & 0.4092 & 0.2835 & 0.0593 & 0.5654 &  \\
    XPY   & 0.9464 & 0.7448 & 0.3610 & 1.1486 &  \\
    XRP   & 0.7563 & 0.3766 & 0.0892 & 0.6966 & * \\
    XTZ   & 0.5616 & 0.6531 & 0.3249 & 1.0066 &  \\
    XUC   & 0.8137 & 0.3398 & 0.0849 & 0.6322 & * \\
    YBC   & 1.0464 & 0.6688 & 0.3183 & 1.0112 & * \\
    ZEC   & 0.5013 & 0.2614 & 0.0540 & 0.5587 &  \\
    ZET   & 1.3383 & 0.6914 & 0.3178 & 1.0888 & * \\
		\end{longtable}
}

\end{document}